\documentclass[%
 reprint,
 amsmath,amssymb,
 aps,
]{revtex4-2}

\usepackage{xcolor}
\usepackage{graphicx}
\usepackage{dcolumn}
\usepackage{bm}
\usepackage{subcaption} 

\newcommand{\bq}{\begin{equation}}
	\newcommand{\eq}{\end{equation}}
\newcommand{\bqn}{\begin{eqnarray}}
	\newcommand{\eqn}{\end{eqnarray}}

\begin{document}

\preprint{APS/123-QED}

\title{What does a regular star look like?}

\author{Yu Liang}
 \email{washy2718@outlook.com}
\affiliation{%
School of Big Data and Artificial Intelligence, Fuyang University of Technology, Anhui 236000, China
}
 
\author{Yuhao Cui}%
\affiliation{%
Department of Astronomy, Xiamen University, Fujian 361005, China
}%

\author{Kai Lin}%
 \email{lk314159@hotmail.com}
\affiliation{%
Universidade Federal de Campina Grande, Campina Grande, PB, 58429-900, Brazil
}%

\author{Sen Guo}%
\affiliation{%
College of Physics and Electronic Engineering, Chongqing Normal University, Chongqing 401331, China
}%

\author{V. H. Satheeshkumar}%
\email{vhsatheeshkumar@gmail.com}
\affiliation{%
 Departamento de F\'{i}sica, Universidade Federal do Estado do Rio de Janeiro (UNIRIO), Rio de Janeiro, RJ 22290-240, Brazil
}%

\author{Yang Huang}%
\affiliation{%
School of Physics and Technology, University of Jinan, Shandong 250022, China
}%

\author{Yang-Yi Sun}%
\affiliation{%
School of Geophysics and Geomatics, China University of Geoscience, Wuhan, Hubei 430074, China
}%

\author{Elcio Abdalla}%
\email{eabdalla@usp.br}
\affiliation{%
Institute of Physics, University of S\~ao Paulo,
05314-970 S\~ao Paulo, Brazil.
}%

\date{\today}

\begin{abstract}
Recently, astronomers discovered unusual Einstein cross images of the galaxy HerS-3, which feature a bright central spot. Motivated by studies of images produced by regular stars, it has been proposed that optical appearances caused by compact stars acting as gravitational lenses may account for this central bright spot. We further suggest that images produced by regular stars exhibit additional characteristics distinct from those of ordinary black holes, such as the possible partial or complete absence of secondary images. These phenomena may serve as favorable observational criteria for identifying regular stars in future searches.

\end{abstract}

\maketitle


\section{Introduction}

Over the past decade, many predictions of general relativity have been confirmed in astronomical observations and laboratory experiments. In 2015, LIGO and Virgo collaborations detected gravitational waves from merging black holes, directly confirming both phenomena \cite{LIGO1, LIGO2, LIGO3, LIGO4, LIGO5, LIGO6, LIGO7}. The Event Horizon Telescope (EHT) released the first image of a black hole in the galaxy M87 \cite{Imagin1,Imagin2,Imagin3,Imagin4,Imagin5,Imagin6} in 2019 and later imaged the black hole Sagittarius A* \cite{Imagin7,Imagin8,Imagin9,Imagin10,Imagin11,Imagin12} at the center of our own galaxy, the Milky Way in 2022. These achievements allowed for direct observation of black holes with accretion disks \cite{Gralla_2019}. Research on gravitational waves and black holes earned Nobel Prizes in Physics in 2017 and 2020, respectively.


Today, scientists widely regard the study of dark matter as the next major frontier in astrophysics. Dark matter is a mysterious form of matter. Its properties and characteristics remain largely unknown. Nevertheless, astronomical observations indicate that dark matter accounts for more than 30\% of the total matter–energy content of the Universe. It plays a pivotal role in cosmic evolution. Existing dark matter models include cold dark matter, warm dark matter, and hot dark matter. Cold dark matter is generally believed to be the dominant component, although warm and hot dark matter may also exist. Dark energy is a peculiar form of energy and may point to still unknown forms of matter-energy in the universe \cite{DEDM,DEDMint1,DEDMint2}. Currently, various experimental programs aim to detect dark matter particles using direct or indirect methods. Observations of the cosmic 21 cm line represent a particularly promising approach. Because dark matter influences the temperature of the cosmic gas and provides the scaffolding for gas condensation, it determines many properties of the cosmic dark ages. Thus, the 21 cm line has become an ideal probe of dark matter. The BINGO experiment \cite{BINGO1,BINGO2,BINGO3} was designed precisely for this purpose. Its successful operation is expected to shed light on the nature and properties of dark matter.


In addition, the study of dark matter through gravitational lensing has become a highly active research direction. Last year, an international collaboration of researchers from the United States, France, Germany, and other countries reported the discovery of an Einstein cross image that differs from the conventional case in that it also exhibits a central bright spot. The team suggested that this central bright spot could be linked to the presence of dark matter because such a feature may result from the gravitational effects of compact dark matter objects along the line of sight \cite{LensEinstein}. If dark matter can undergo gravitational condensation, the Universe may host star-like objects composed of dark matter. Since dark matter is expected to have negligible electromagnetic interactions, such dark matter stars could be optically transparent, providing a possible explanation for the central bright spot observed in the Einstein cross.


On the other hand, the problem of black hole singularities has long remained a major challenge in theoretical physics. Although the cosmic censorship conjecture postulates that singularities are separated from our spacetime by a one-way event horizon, fundamental physical difficulties persist. In 1968, Bardeen \cite{Regularblackhole1} proposed a novel approach by constructing a black hole metric in which the central singularity is removed; such objects are known as regular black holes. Subsequently, various forms of regular black hole line elements have been proposed under different assumptions and constructions: see for example works by Dymnikova \cite{Regularspacetime2,Regularspacetime3,Regularspacetime4} and Hayward \cite{Regularblackhole2}. As pointed out in Ref. \cite{RegularStar}, if the parameters of a regular black hole are further increased, the event horizon may disappear, leading to a horizonless spacetime without a central singularity, referred to as a regular star. If a regular star is considered as a type of dark-matter star object, an important question arises: can it account for the central bright spot observed in the Einstein cross reported in Ref. \cite{LensEinstein}? This study addresses this question.


This work is organized as follows. In Sect. II, we use the Bardeen \cite{Regularblackhole1} and Hayward \cite{Regularblackhole2} spacetimes as representative examples. Using these, we introduce several typical regular stars and derive the equations governing the geodesic motion of particles and light in these spacetime backgrounds. In Sect. III, we present the photon trajectories and images in these spacetimes. We discuss the image properties of the star under different conditions. In this section, we first investigate the images of stars surrounded by geometrically and optically thin accretion disks. We then demonstrate the images formed by gravitational lensing through the stellar center. Next, we study the geometrically thin and optically thick accretion disk model. Here, we emphasize the incomplete secondary images unique to regular stars, which do not appear in black hole spacetimes. Such image features may provide a potential observational signature for distinguishing black holes from regular stars in the future. Sect. IV contains our conclusions and discussions.

\section{Null Geodesics in regular star spacetime}

Here we derive the equations governing the geodesic motion of particles and light in spacetime backgrounds without  singularity and event horizon. In Ref. \cite{RegularStar}, we showed that for traditional regular black holes, when the parameter $\beta$ responsible for the removal of the singularity becomes sufficiently large, the event horizon disappears. However, since there is no singularity at the center of a regular black hole, the resulting spacetime does not contain a naked singularity and thus represents a "legitimate" star spacetime.


A well-known regular spacetime is the Hayward star spacetime, whose metric can be written as\cite{Regularblackhole1,Regularblackhole2}:
\begin{equation}
    d s^2 = -f dt^2 + \frac{1}{f} dr^2 + r^2 d\theta^2 + r^2 \sin^2\theta d\phi^2
\end{equation}
where the Hayward case and the Bardeen case, respectively, are given by
\begin{align}
    f = f_{\rm Hayward}(r,\beta) = 1 - \frac{2 M r^2}{r^3 + 2\beta^2} \label{eq:fr_hayward} \\
    f = f_{\rm Bardeen}(r,\beta) = 1 - \frac{2 M r^2}{(r^2 + \beta^2)^{3/2}}\label{eq:fr_bardeen}
\end{align}

An analysis shows that for different values of $\beta$, the metric describes distinct spacetime structures. When $\beta < \beta_e$, the spacetime corresponds to a black hole, and the root of $f=0$ defines the radius of the event horizon $r_e$, which is determined by the equation $f=0$. When $\beta_e \leq \beta < \beta_c$, the spacetime describes a configuration with a photon sphere. When $\beta \geq \beta_c$, the spacetime corresponds to a configuration without a photon sphere. For Hayward case, the explicit expressions for $\beta_e$ and $\beta_c$ are given by
\begin{equation}
    \beta_e = \frac{4}{3} \sqrt{\frac{M^3}{3}}, \qquad
    \beta_c = \frac{25}{24} \sqrt{\frac{5 M^3}{6}}.
\end{equation}
For the Bardeen case, the explicit expressions for $\beta_e$ and $\beta_c$ are given by
\begin{equation}
    \beta_e = \frac{4M}{3\sqrt{3}}, \qquad
    \textcolor{black}{\beta_c = \frac{48M}{25 \sqrt{5}}.}
\end{equation}

It is evident that, for any spherically symmetric metric, the equations of motion for photons, derived from the Hamiltonian, take the following form:
\begin{equation}\label{eq:move}
    \left[ \frac{1}{r^2} \left( \frac{dr}{d\phi} \right) \right]^2 + \frac{f}{r^2} = \frac{1}{b^2}
\end{equation}
where $b=L/E$ is the impact parameter.


For cases where there is no black hole or where a black hole does not capture photons, the photon trajectory has a periastron $r_n$ satisfying $d r_n / d \phi = 0$. From the above equation, the relationship between the impact parameter and the periastron is given by Eq.(\ref{eq:b_rn}). When the impact parameter $b$ is known, the corresponding periastron $r_n$ can be numerically solved from this equation. Therefore,
\begin{equation}\label{eq:b_rn}
    b = \frac{r_n}{\sqrt{f(r_n,\beta)}}
\end{equation}


For the case where a photon sphere $r_p$ exists, the additional condition $d^2 r / d \phi^2 = 0$ is satisfied. The corresponding impact parameter in this case is the critical impact parameter $b_c$. Solving these equations simultaneously, one finds that the following relation holds when $r=r_p$:
\begin{equation}
    f - \frac{r}{2} \frac{\partial f}{\partial r} = 0
\end{equation}

We define $r=1/u$, so that the equation of motion~(\ref{eq:move}) can be rewritten in the following form:
\begin{equation}
    \left( \frac{du}{d\phi} \right)^2 = \frac{1}{b^2} - u^2 f_u \equiv G(u)
\end{equation}
where $f_u$ is given by Eq.(\ref{eq:fr_hayward}) or Eq.(\ref{eq:fr_bardeen}).

\begin{figure*}[htbp]  
    \centering
    \begin{subfigure}{0.19\textwidth}  
        \centering
        \includegraphics[width=\linewidth]{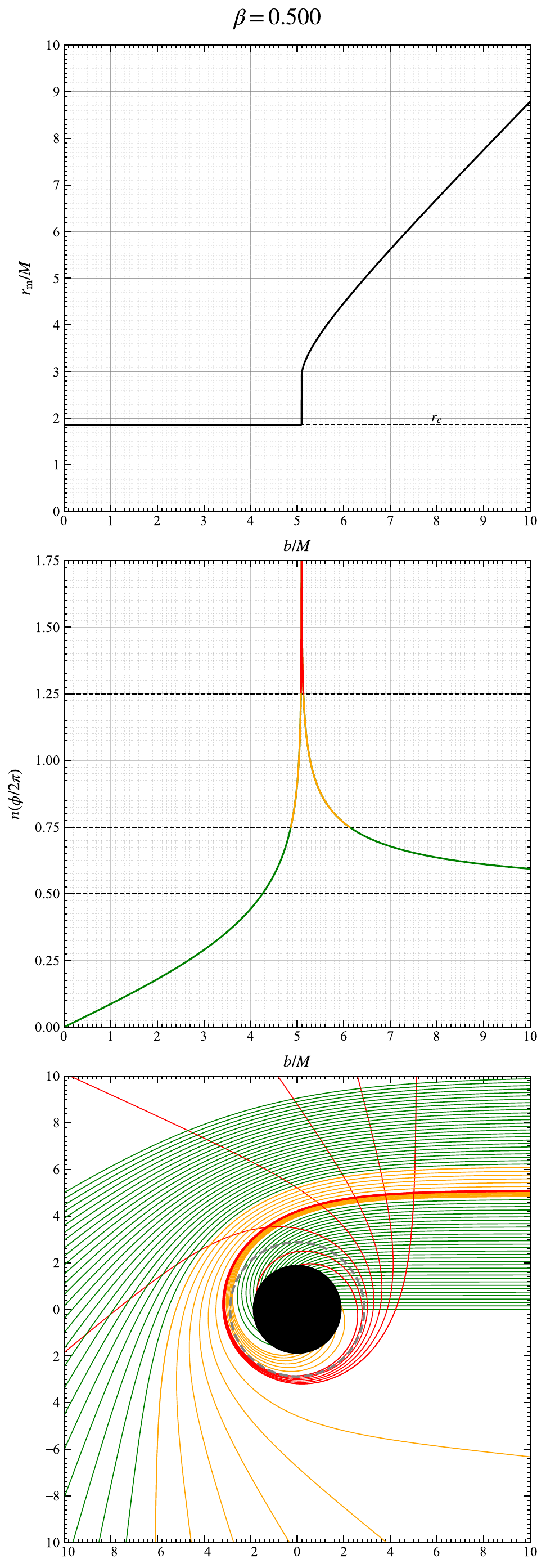}  
    \end{subfigure}
    \hfill  
    \begin{subfigure}{0.19\textwidth}
        \centering
        \includegraphics[width=\linewidth]{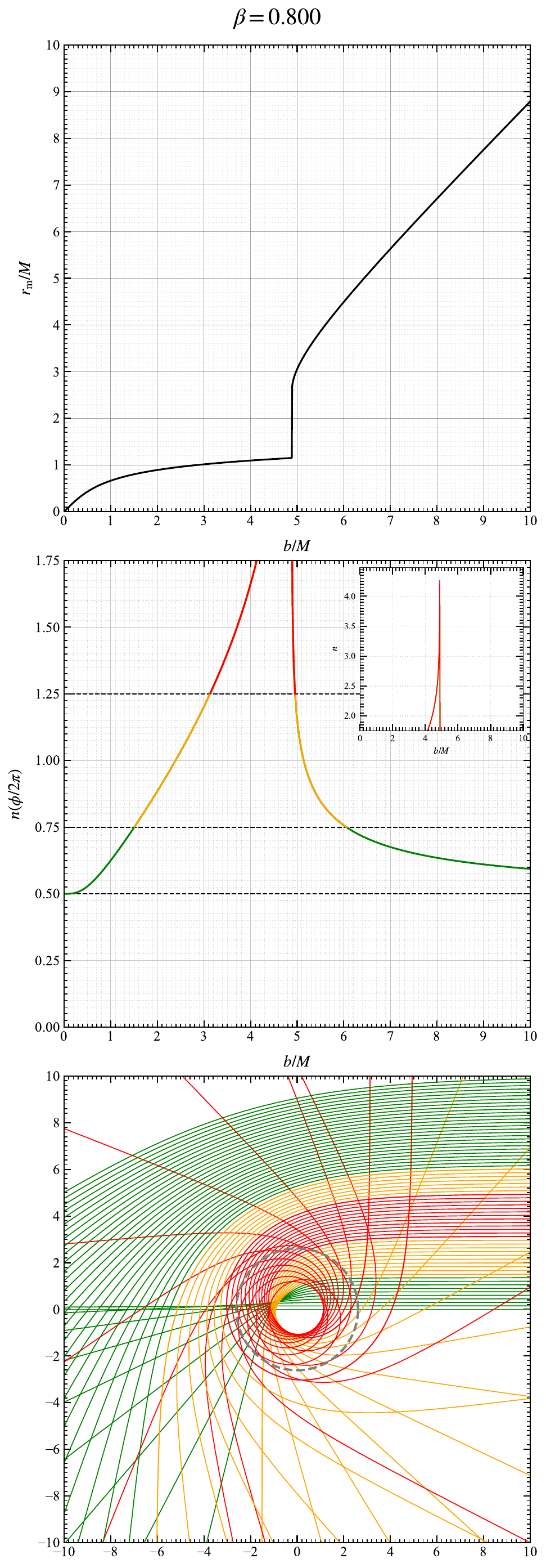}
    \end{subfigure}
    \hfill
    \begin{subfigure}{0.19\textwidth}
        \centering
        \includegraphics[width=\linewidth]{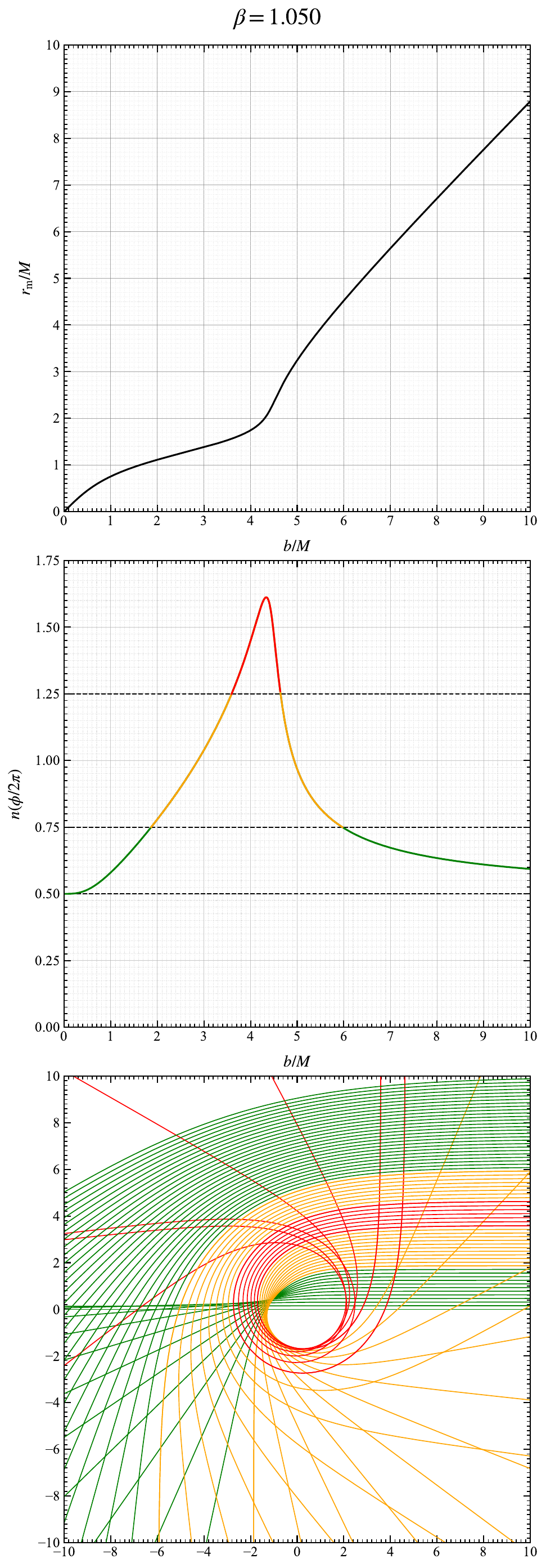}
    \end{subfigure}
    \hfill
    \begin{subfigure}{0.19\textwidth}
        \centering
        \includegraphics[width=\linewidth]{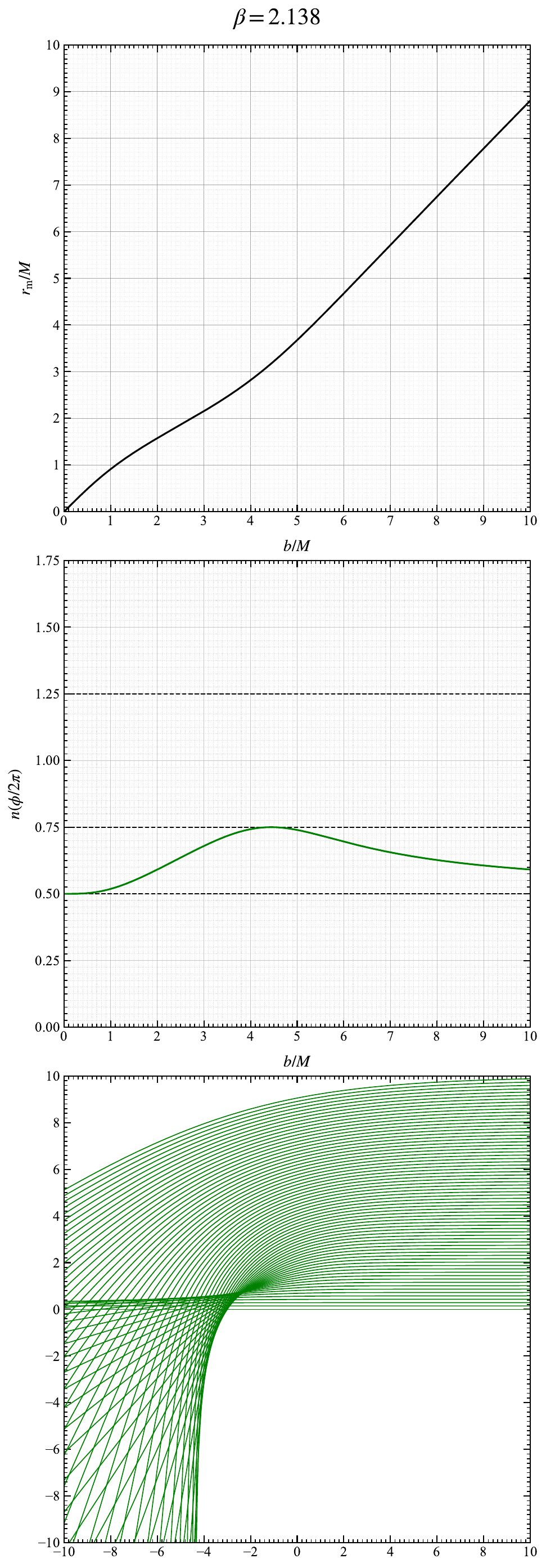}
    \end{subfigure}
    \hfill
    \begin{subfigure}{0.19\textwidth}
        \centering
        \includegraphics[width=\linewidth]{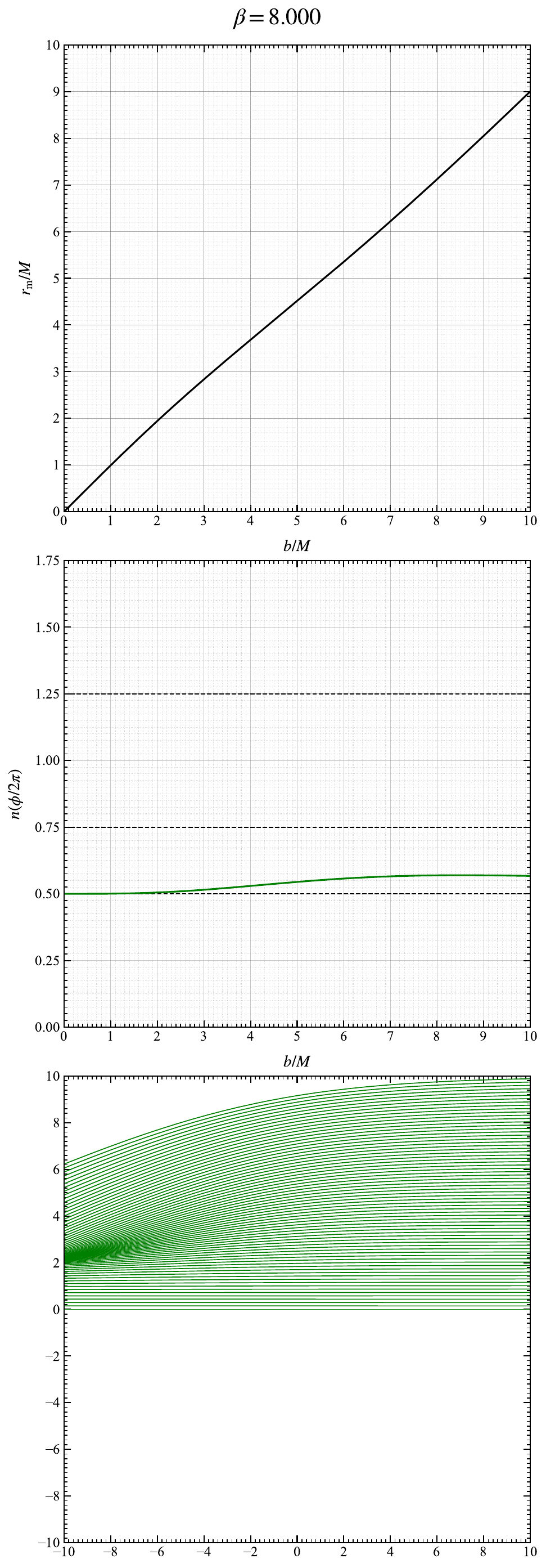}
    \end{subfigure}
    \caption{Null geodesics in the Hayward spacetime for different $\beta$ values, where the first row presents $r_m$ as a function of the impact parameter $b$, the second row presents the orbit number $n$ versus $b$ with green, orange, and red indicating \textcolor{black}{the direct, lensed, and photon ring}, respectively, and the third row displays the light trajectories for $b \in (0,10)$ using the same color scheme as the second row.}  
    \label{fig:light_ray}  
\end{figure*}

We define the deflection angle of a photon traveling from infinity to its closest approach to the black hole as the maximal deflection angle $\phi_{\max}$. In a photon trajectory, the closest point to the black hole is usually either the event horizon or the periastron, and these two cases are classified as in Eq.(\ref{eq:rm}). The first row of Fig.\ref{fig:light_ray} shows the relation between $r_m$ and $b$ for different values of $\beta$.

\begin{equation}
\label{eq:rm}
    r_m = \left\{
    \begin{split}
        &r_e, \quad && \beta < \beta_e {\rm~~and~~} b<b_c.\\
        &r_n, \quad && \beta \geq \beta_e.
    \end{split}
    \right.
\end{equation}
Therefore, for any photon trajectory, the equation for calculating the maximal deflection angle is given by:
\begin{equation}
    \phi_{\max} = \int_0^{\frac{1}{r_m}} \frac{du}{\sqrt{G(u)}}
\end{equation}
For photons with $r_m = r_n$, the trajectory also includes the outgoing portion after passing the periastron, which is symmetric with respect to the incoming portion about the periastron. According to the definition of the number of orbits $n \equiv \phi / 2\pi$ in Reference \cite{Gralla_2019} and the above equation, the corresponding formula for $n$ is given by Eq.(\ref{eq:n}). The second row of Fig.\ref{fig:light_ray} shows the relation between $n$ and $b$ for different values of $\beta$.

\begin{equation}\label{eq:n}
    n = \left\{
    \begin{split}
        &\phi_{\max}, \quad && r_m=r_e, \\
        &2 \phi_{\max}, \quad && r_m=r_n.
    \end{split}
    \right.
\end{equation}


When we need to calculate the complete trajectory of a photon, it is necessary to divide it into the portions before and after passing $r_m$ (if the latter exists). The portion before passing $r_m$ can be calculated using the following formula ($r \in (r_m,+\infty)$):
\begin{equation}
    \phi(r) = \int_0^{\frac{1}{r}} \frac{du}{\sqrt{G(u)}}
\end{equation}
The portion after passing $r_m$ can be calculated using the following formula ($r \in (r_m,+\infty)$):
\begin{equation}
    \phi(r) = 2 \phi_{\max} - \int_0^{\frac{1}{r}} \frac{du}{\sqrt{G(u)}}
\end{equation}

The third row of Fig.\ref{fig:light_ray} shows the photon trajectories for different impact parameters. It can be seen that when $\beta>\beta_e$, since there is no singularity in the spacetime structure, the photon trajectory always possesses a periastron, and the photon emerges after passing the periastron. If dark matter has a similar spacetime structure and is ``transparent", a central bright spot of an Einstein cross could potentially appear.

\begin{figure*}[htbp]  
    \centering
    \begin{subfigure}{0.19\textwidth}  
        \centering
        \includegraphics[width=\linewidth]{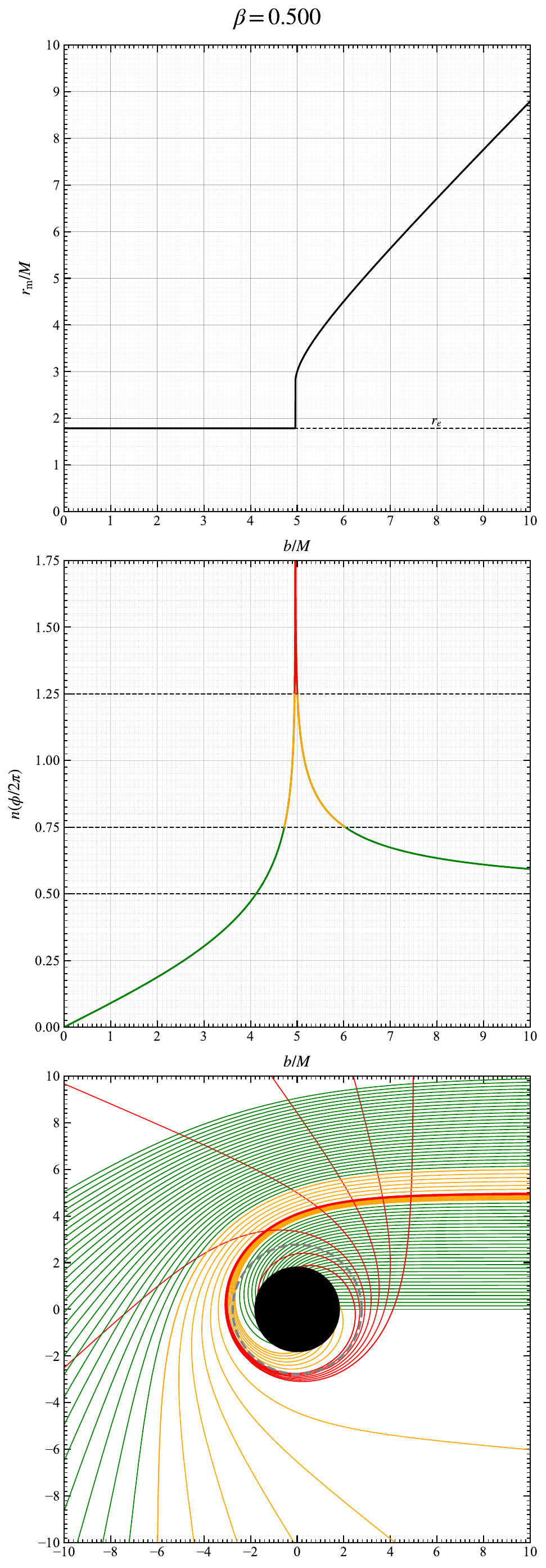}  
    \end{subfigure}
    \hfill  
    \begin{subfigure}{0.19\textwidth}
        \centering
        \includegraphics[width=\linewidth]{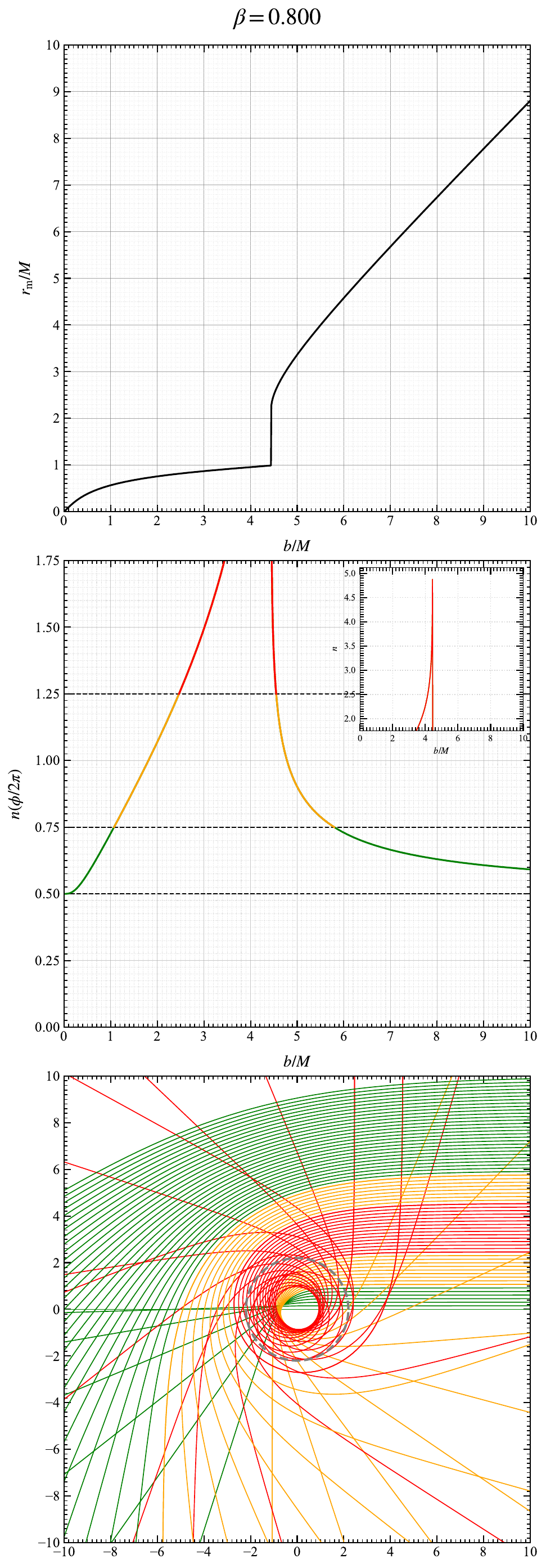}
    \end{subfigure}
    \hfill
    \begin{subfigure}{0.19\textwidth}
        \centering
        \includegraphics[width=\linewidth]{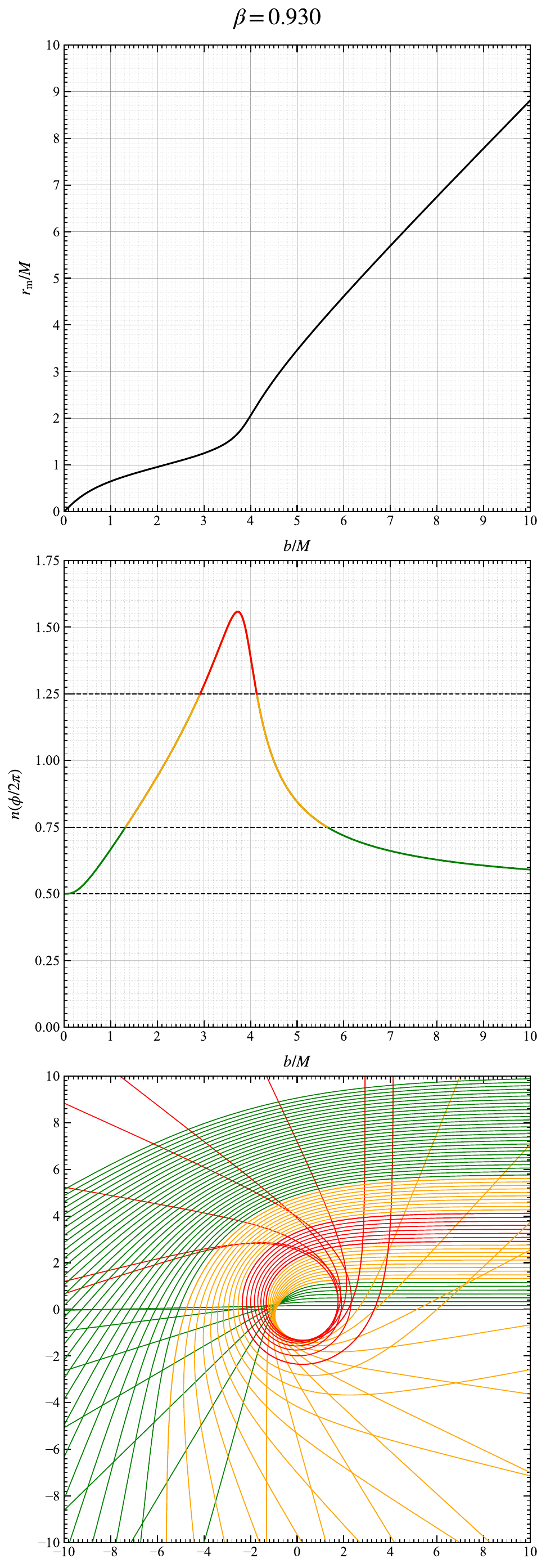}
    \end{subfigure}
    \hfill
    \begin{subfigure}{0.19\textwidth}
        \centering
        \includegraphics[width=\linewidth]{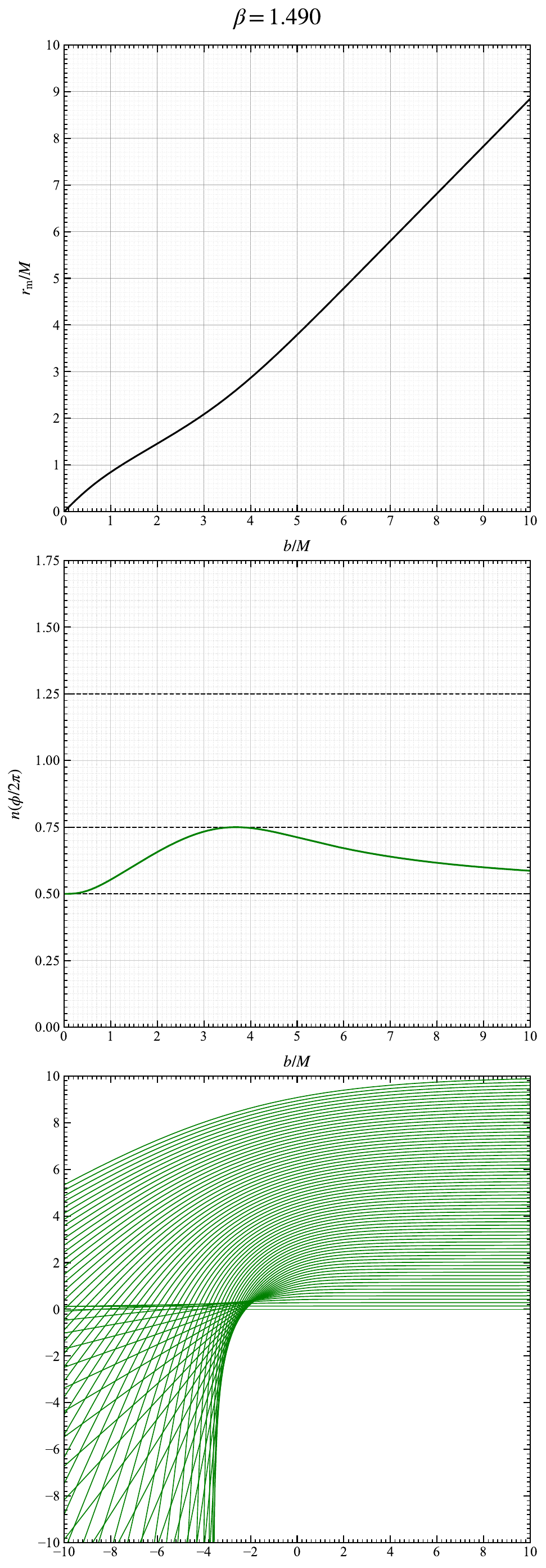}
    \end{subfigure}
    \hfill
    \begin{subfigure}{0.19\textwidth}
        \centering
        \includegraphics[width=\linewidth]{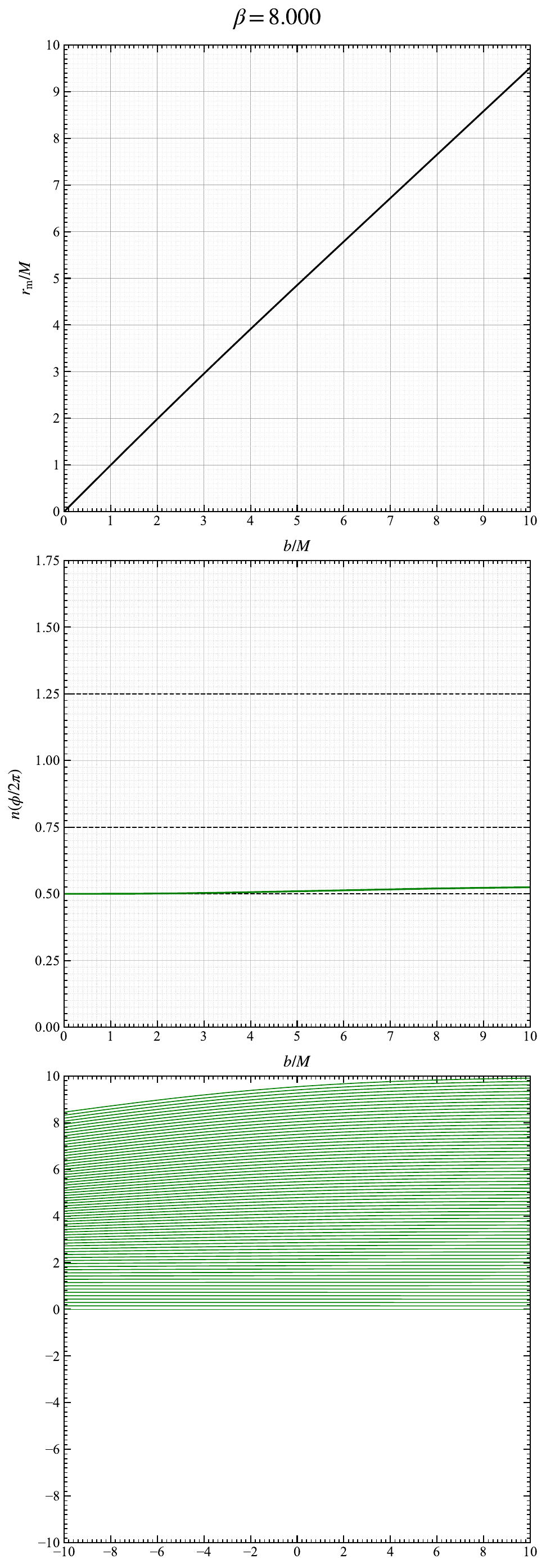}
    \end{subfigure}
    \caption{Null geodesics in the Bardeen spacetime for various $\beta$ values, analogous to those shown in Fig.\ref{fig:light_ray}.}  
    \label{fig:bardeen_light_ray}  
\end{figure*}

Fig.\ref{fig:bardeen_light_ray} shows the photon trajectories for the Bardeen metric with different values of $\beta$. The conclusions are similar to those for the Hayward metric. 
{Since the two metrics share similar properties, only the results for the Hayward metric will be presented in the following.}

When $\beta$ becomes sufficiently large, the regular spacetime has no event horizon, so the region near the center of the star can be transparent. This feature is the most significant distinction from a black hole. Although the event horizon disappears, the photon sphere of the object may still exist. Photons approaching at certain angles can orbit the photon sphere multiple times. However, unlike a black hole, photons entering the interior of the photon sphere experience a weaker gravitational pull, resulting in fewer orbits. This can be equivalently interpreted as a reduction of the effective gravitational force inside the photon sphere. This behavior is similar to the conclusions of the shell theorem in Newtonian gravity—for example, an object inside the Earth experiences a gravitational force smaller than that at the surface, and at the center of the Earth, the net gravitational force can even be zero. This is also a manifestation of the symmetry of a spherically symmetric object. If $\beta$ increases further, the photon sphere of the star may disappear, meaning that in such a spacetime, photons cannot orbit near the star multiple times. When $\beta$ exceeds a certain critical value, photons in this spacetime can only be slightly deflected by gravity, and the spacetime can only produce weak gravitational lensing.

Next, we take $\beta = 5$ as an illustrative example to show the gravitational lensing images produced in a regular star spacetime, in which images appear both at the center and in the outer regions.

\begin{figure*}
    \centering
    \includegraphics[width=\linewidth]{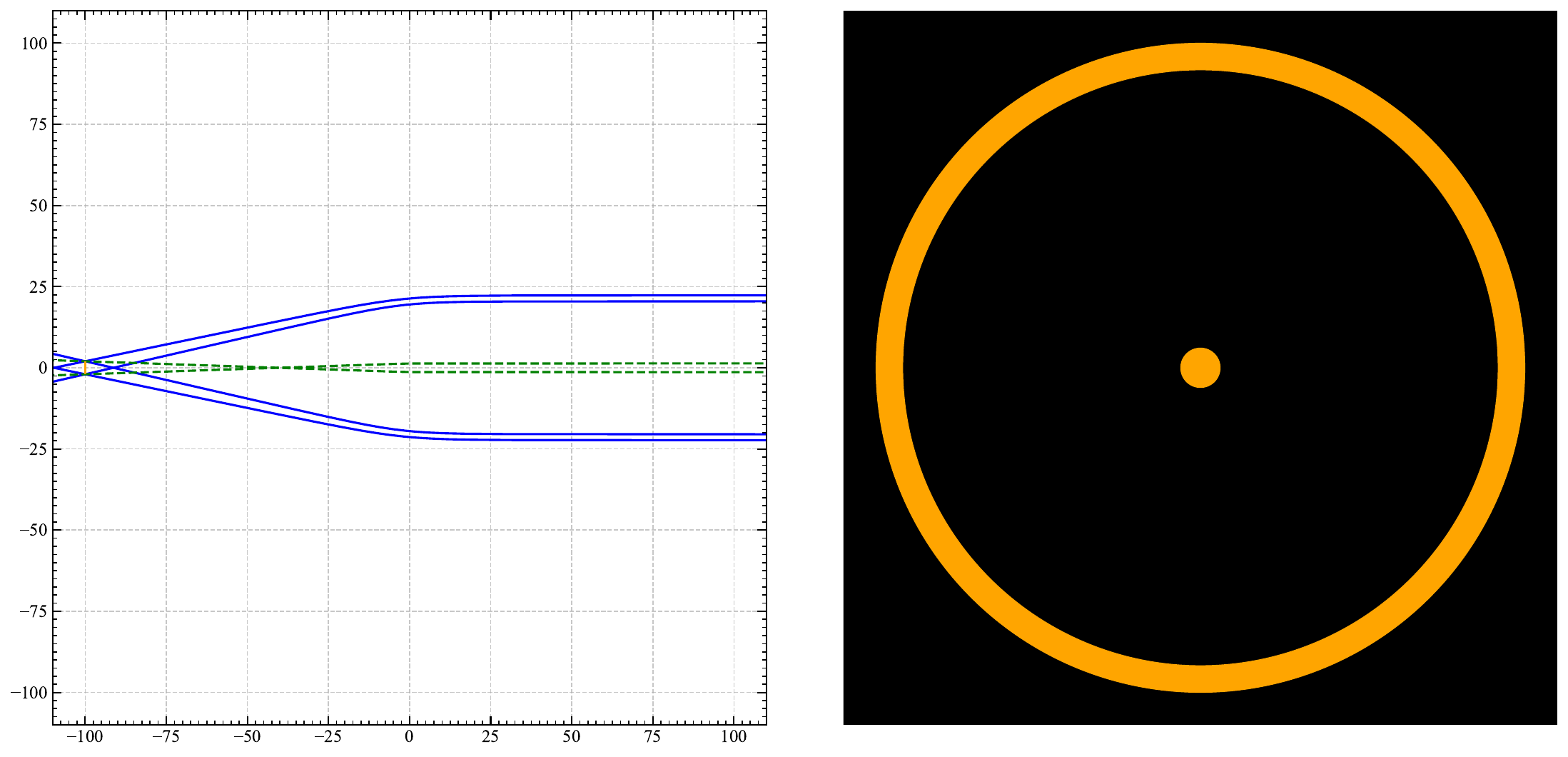}
    \caption{\textcolor{black}{Gravitational lensing induced by a transparent spacetime configuration described by the Hayward metric with $\beta=5$: the left panel shows an emitting disk of radius $2M$ located at $-100M$ (orange solid line), light rays propagating around the exterior of the spacetime structure (blue solid lines), and light rays traversing the interior of the structure (green solid lines), with the observer plane placed at $+\infty$; the corresponding lensed image on the observer plane is displayed in the right panel.}}
    \label{fig:e_rays_image}
\end{figure*}

We consider a transparent spacetime structure described by the Hayward metric with $\beta=5$ located between a distant luminous source and the observer, assuming that the source is positioned at $-100M$ with a radius of $2M$, the spacetime structure is centered at the coordinate origin, and the observer plane is located at $+\infty$. Here, the three-dimensional structure of the emitter is ignored and approximated as a disk, which allows us to avoid more involved geometric calculations that do not significantly affect the results. Based on ray-tracing calculations, we plot the light trajectories under gravitational lensing effects, as shown in Fig.~\ref{fig:e_rays_image}. In this figure, blue rays denote light trajectories that circumvent the spacetime structure externally, whereas green rays indicate trajectories in which light propagates directly through the structure. On the observer plane, a luminous ring and a central bright spot are observed, as shown in Fig.~\ref{fig:e_rays_image}.

In the next section, we present the images of the star for different accretion disk models.

\section{Radiation flux}


In this work, we consider two typical accretion disk models, namely the geometrically and optically thin disk and the geometrically thin and optically thick disk.


In order to better analyze the radiation intensity, we first need to calculate the Innermost Stable Circular Orbit (ISCO). The ISCO must simultaneously satisfy the circular orbit conditions $\dot{r} = 0$ and $\ddot{r} = 0$, as well as the second derivative of the effective potential $V" (r) = 0$. The effective potential $V(r)$ is defined as
\begin{equation}
    V(r) = f \left( \frac{L^2}{r^2} + 1 \right)
\end{equation}
From our derivation, the ISCO of a spherically symmetric black hole is given by the root of the following equation:
\begin{equation}
    f (f'+r f'')+2r f'^2 = 0
\end{equation}

From the calculation of the above equation, it can be seen that when $\beta < \beta_e$, there exists only one valid real root. When $\beta \geq \beta_e$, there are two valid real roots, where the larger root corresponds to the ISCO, hereafter denoted as $r_+$, and the smaller root corresponds to the outer boundary of the inner stable circular orbit, hereafter denoted as $r_-$.

\subsection{Geometrically and Optically Thin Disk Emission}

\begin{figure*}[htbp]  
    \centering
    \begin{subfigure}{0.19\textwidth}  
        \centering
        \includegraphics[width=\linewidth]{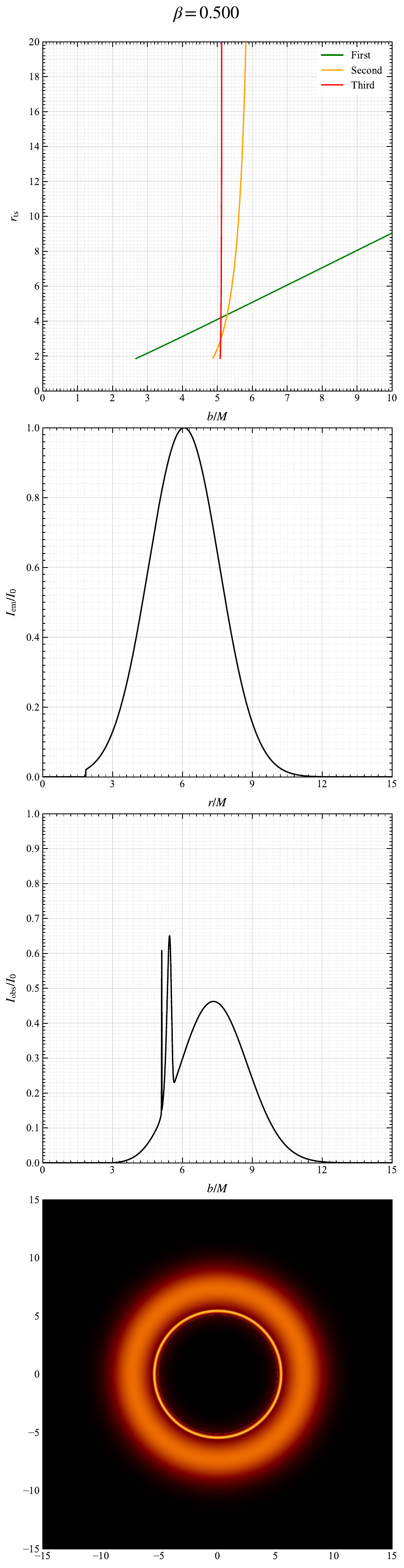}
    \end{subfigure}
    \hfill  
    \begin{subfigure}{0.19\textwidth}
        \centering
        \includegraphics[width=\linewidth]{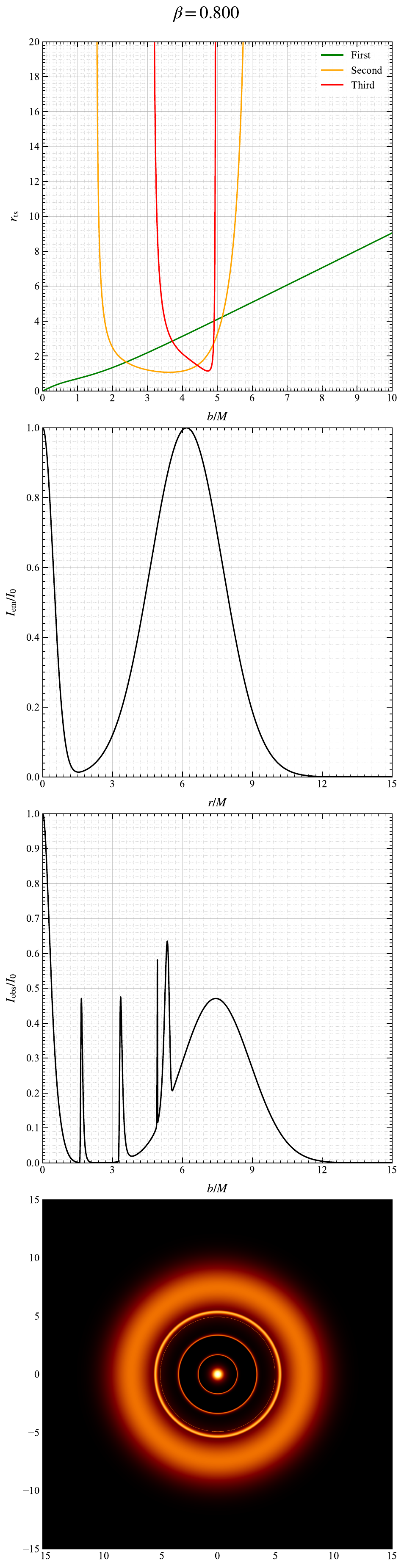}
    \end{subfigure}
    \hfill
    \begin{subfigure}{0.19\textwidth}
        \centering
        \includegraphics[width=\linewidth]{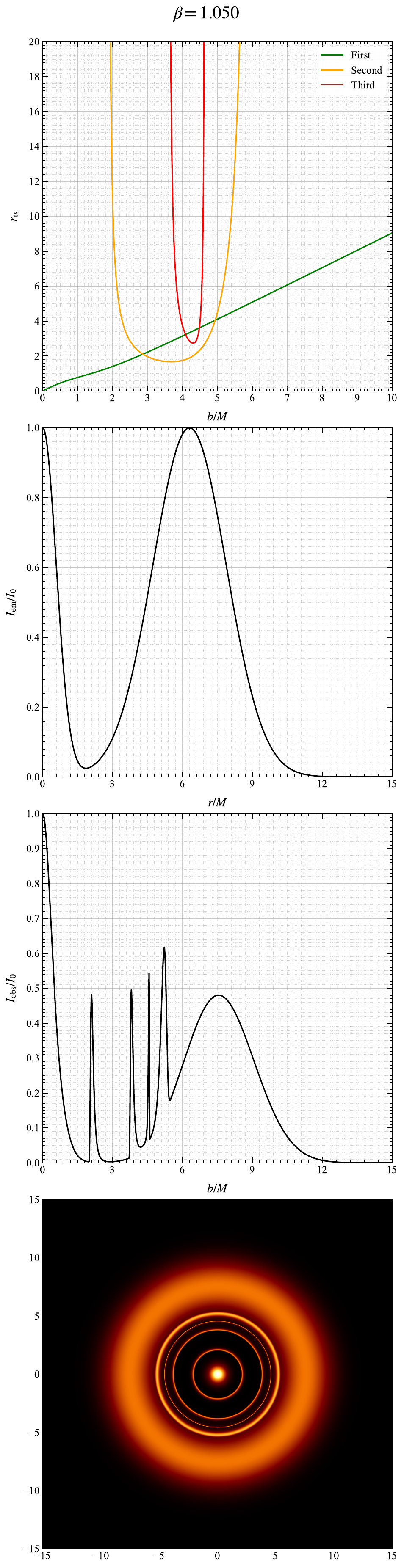}
    \end{subfigure}
    \hfill
    \begin{subfigure}{0.19\textwidth}
        \centering
        \includegraphics[width=\linewidth]{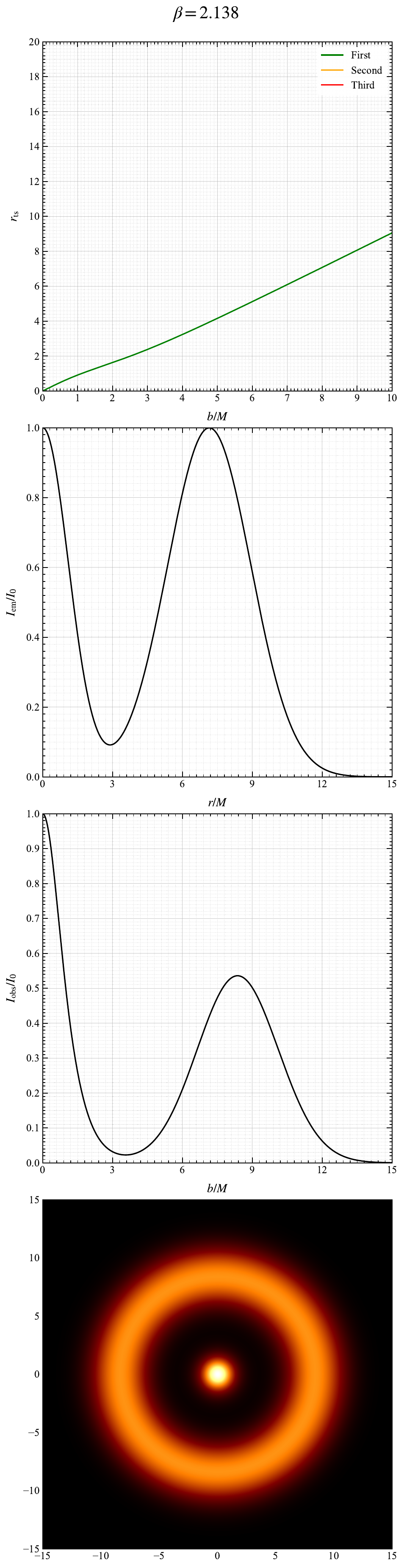}
    \end{subfigure}
    \begin{subfigure}{0.19\textwidth}
        \centering
        \includegraphics[width=\linewidth]{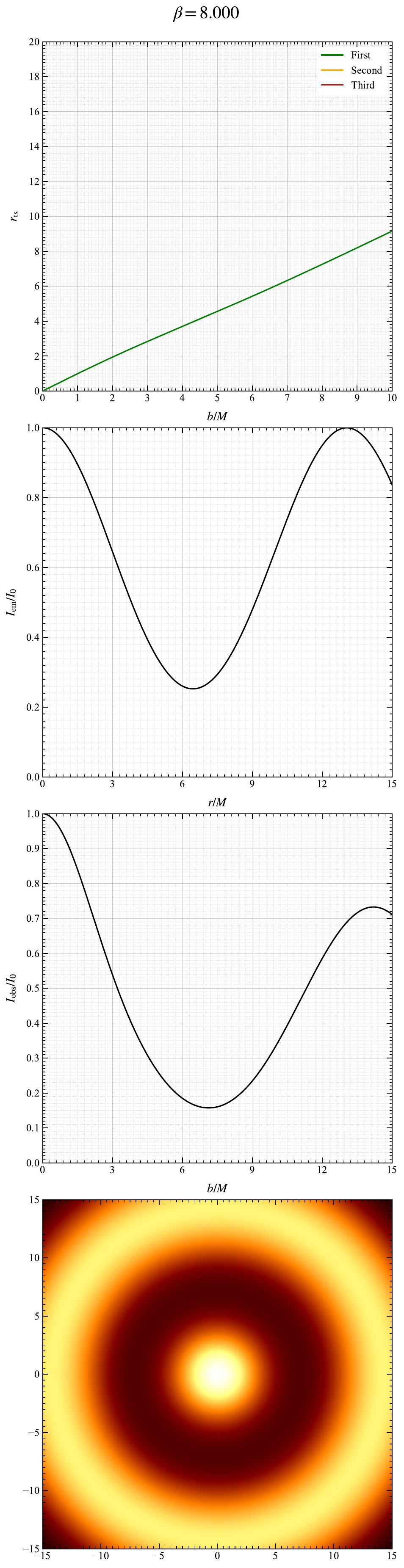}
    \end{subfigure}
    \caption{Images of a geometrically and optically thin accretion disk for different values of $\beta$. The first row shows the dependence of the transfer function $r_{ts}$ on the impact parameter $b$, where green, orange, and red denote the distances to the first three intersection points between the light ray and the disk plane. The second row shows the radial dependence of the normalized emission intensity $I_{em}$. The third row displays the normalized observed intensity $I_{obs}$ as a function of $r$. The fourth row presents the resulting images of the geometrically and optically thin accretion disk.}  
    \label{fig:thin_image}
\end{figure*}

The accretion-disk model adopted here draws on the studies of Gralla and collaborators concerning black hole images \cite{Gralla_2019, Gao_2023mjb}. An interesting feature of the current metric is that when $\beta \geq \beta_e$, there is no singularity in the spacetime structure, and matter can be stably distributed within the range $r \in (0,r_-) \cup (r_+,+\infty)$. Accordingly, we construct a toy model for the radiation intensity as shown in Eq.(\ref{eq:Iem}) to study the observational images in this spacetime. The second row of Fig.\ref{fig:thin_image} shows the variation of the radiation intensity $I_\text{em}$ with the radial distance $r$ for different values of $\beta$, normalized by the maximal radiation intensity. $I_\text{em}$ is given by
\begin{equation}\label{eq:Iem}
\textcolor{black}{
    I_\text{em} = \begin{cases}
        e^{-8(r-r_+)^2/r_+^2}, &\beta<\beta_e\ {\rm and}\ r > r_e\\
        e^{-r^2/(2 r_-)} + e^{-8(r-r_+)^2/r_+^2}, &\beta \geq \beta_e\ {\rm and}\ r > 0
    \end{cases}
}
\end{equation}

When tracing photon trajectories, Gralla defined a transfer function $r_{ts}(b)$ to describe the radial distance at which a photon with the impact parameter $b$ intersects the disk \cite{Gralla_2019}. It is easy to see that the intersection distances for photons with different impact parameters vary significantly. The first row of Fig.\ref{fig:thin_image} shows the relation between $r_{ts}$ and $b$ for different values of $\beta$.


From the transfer function, the observed intensity $I_\text{obs}$ of any photon trajectory can be expressed as:
\begin{equation}
    I_\text{obs}(b) = \sum_n g^4 I_\text{em}\vert_{r=r_{ts}(b)}
\end{equation}
Here, $g = \sqrt{f}$ accounts for the effect of gravitational redshift. The third row of Fig.\ref{fig:thin_image} shows the observed intensity images for different values of $\beta$. It can be seen that for $\beta > 0$, there is both a prominent multi-ring structure and a distinct central bright spot. The number of secondary rings is related to the number of intersections with the disk—the more intersections, the more secondary rings appear. The central bright spot arises because this spacetime structure has no singularity, and the central region can stably contain radiating matter, producing the bright spot. Here, we assume that the entire spacetime structure is transparent.

\subsection{Geometrically Thin and Optically Thick Disk Emission}

\begin{figure}
    \centering
    \includegraphics[width=\linewidth]{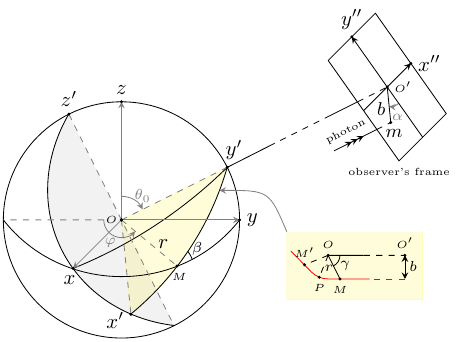}
    \caption{The coordinate system \cite{luminet_1979,tian_2019}.}
    \label{fig:tian_2019_fig3}
    
\end{figure}

We also present the images obtained under the geometrically thin and optically thick disk model. The ray-tracing treatment of this accretion-disk model follows that of standard black hole accretion-disk models \cite{luminet_1979,tian_2019,zhang_2021,hou_2022,meng_2023,guo_2024}. The coordinate system is constructed as illustrated in Fig.\ref{fig:tian_2019_fig3}, where the deflection angle $\gamma$ obeys the following mathematical relationship with the observer's inclination angle $\theta_0$ and $\alpha$ \cite{luminet_1979,tian_2019}
\begin{equation}
    \cos \gamma = \frac{\cos \alpha}{\sqrt{\cos^2 \alpha + \cot^2 \theta_0}}
\end{equation}

\begin{figure}
    \centering
    \includegraphics[width=\linewidth]{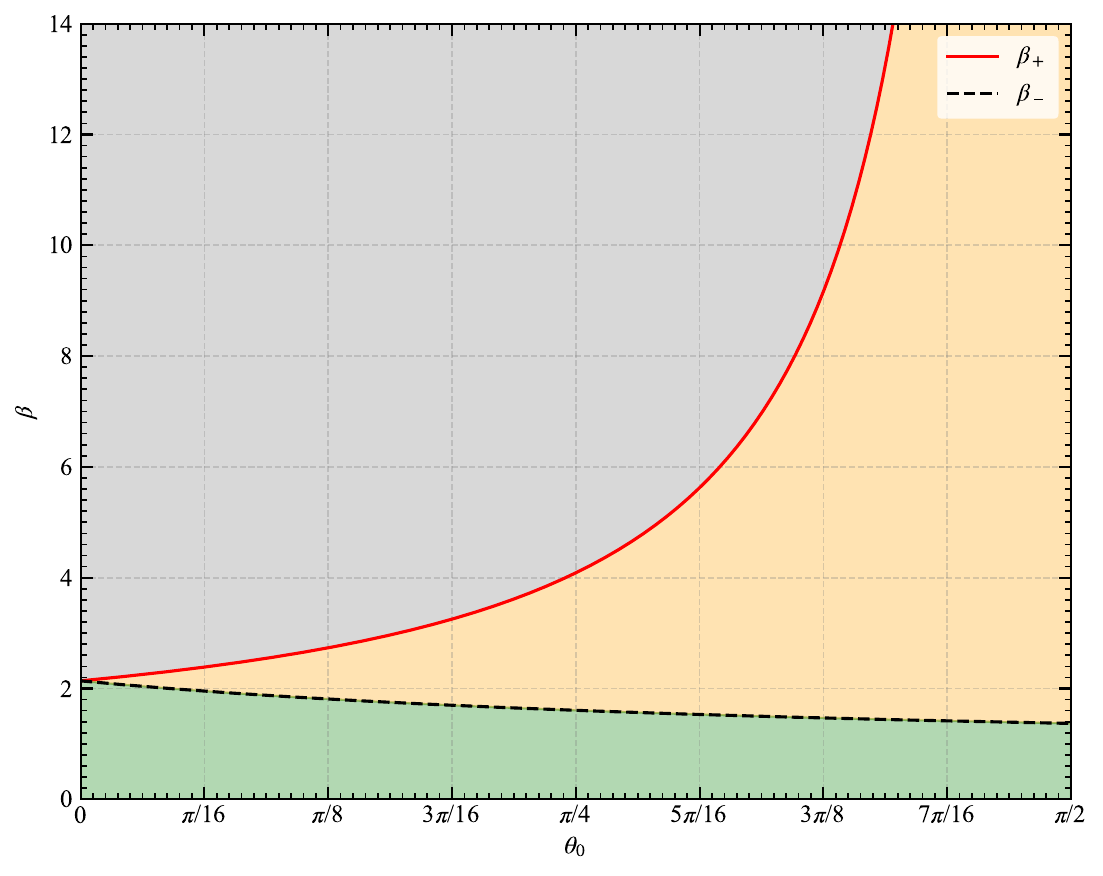}
    \caption{The dependence of the secondary image on $\beta$ and $\theta_0$. The green area corresponds to cases where the secondary image is fully observable, the orange area represents incomplete secondary images, and the gray area signifies the nonexistence of the secondary image.}
    \label{fig:theta0_beta}
\end{figure}

For such regular spacetimes, the secondary images can become incomplete or even disappear, and this behavior is jointly influenced by $\beta$ and the observer inclination angle $\theta_0$. For a fixed $\theta_0$, we define $\beta=\beta_-$ as the case in which the secondary image is just fully visible, and $\beta=\beta_+$ as the case in which the secondary image is just completely invisible. The condition for $\beta_-$ is that the maximum orbit number satisfies $n_{\max}\le 3\pi/2-\theta_0$, while the condition for $\beta_+$ is $n_{\max}\ge 3\pi/2+\theta_0$. Fig~\ref{fig:theta0_beta} shows the dependence of $\beta_\pm$ on $\theta_0$, where the green region represents cases in which the secondary image is fully visible, the orange region indicates incomplete secondary images, and the gray region signifies that no secondary image is present.

The corresponding expression for the radiation intensity is given by \cite{page_1974}
\begin{equation}
    I_{em} = - \frac{\dot{M_0}}{4 \phi \sqrt{-g}} \frac{\Omega_{,r}}{(E - \Omega L)^2} \int_{r_+}^{r} (E - \Omega L) L_{,r} dr
\end{equation}
where $\dot{M_0}$ is the mass accretion rate, $g$ is the metric determinant. $E$, $L$, and $\Omega$ represent the energy, angular momentum, and angular velocity, and are given by
\begin{equation}
    E = - \frac{g_{tt}}{\sqrt{-g_{tt} - g_{\phi\phi}\Omega^2}}
\end{equation}
\begin{equation}
    L = \frac{g_{\phi\phi} \Omega}{\sqrt{-g_{tt} - g_{\phi\phi}\Omega^2}}
\end{equation}
\begin{equation}
    \Omega = \frac{d\phi}{dt} = \sqrt{-\frac{g_{tt,R}}{g_{\phi\phi,R}}}
\end{equation}
For a thin accretion disk confined to the equatorial plane, the corresponding components are given by
\begin{equation}
    g_{tt} = -f, \quad
    g_{\phi\phi} = r^2, \quad
    g = -r^4.
\end{equation}

According to Ref.~\cite{luminet_1979}, the observed flux can be expressed as
\begin{equation}
    I_{obs} = \frac{I_{em}}{(1+z)^4}
\end{equation}
where $(1 + z)$ denotes the redshift factor, given by
\begin{equation}
    1 + z = \frac{1 + b \Omega \sin \theta_0 \sin \alpha}{\sqrt{-g_{tt}-g_{\phi\phi} \Omega^2}}
\end{equation}

\begin{figure*}[htbp]  
    \centering
    \begin{subfigure}{0.19\textwidth}  
        \centering
        \includegraphics[width=\linewidth]{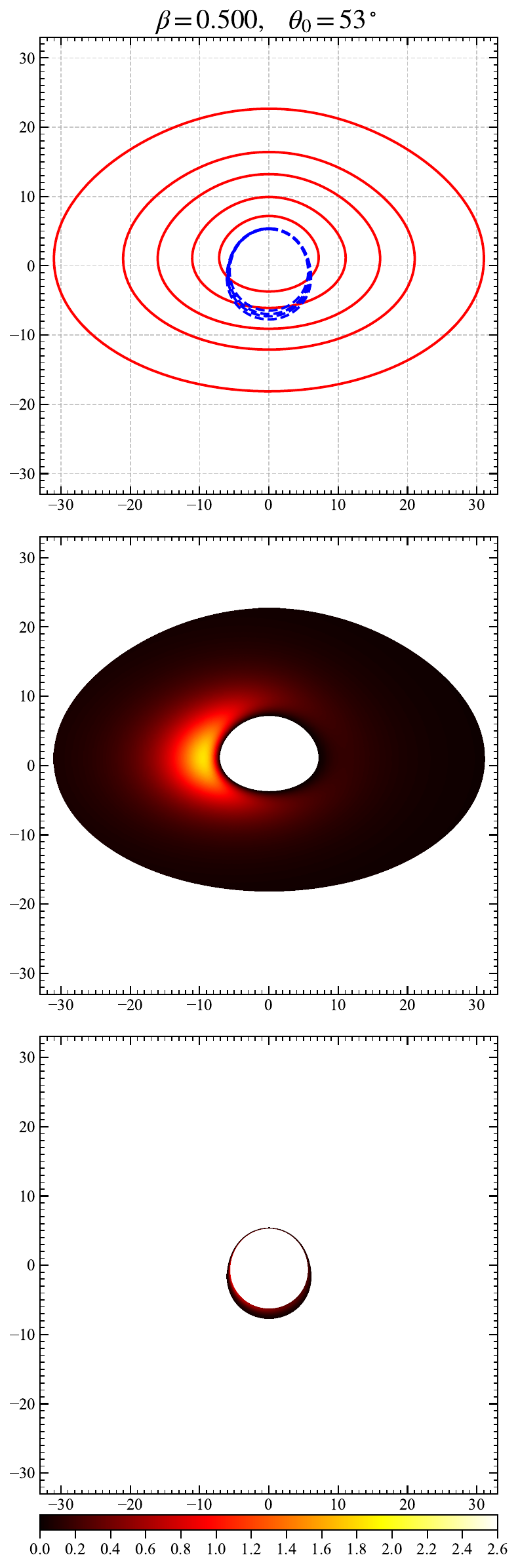}
    \end{subfigure}
    \hfill  
    \begin{subfigure}{0.19\textwidth}
        \centering
        \includegraphics[width=\linewidth]{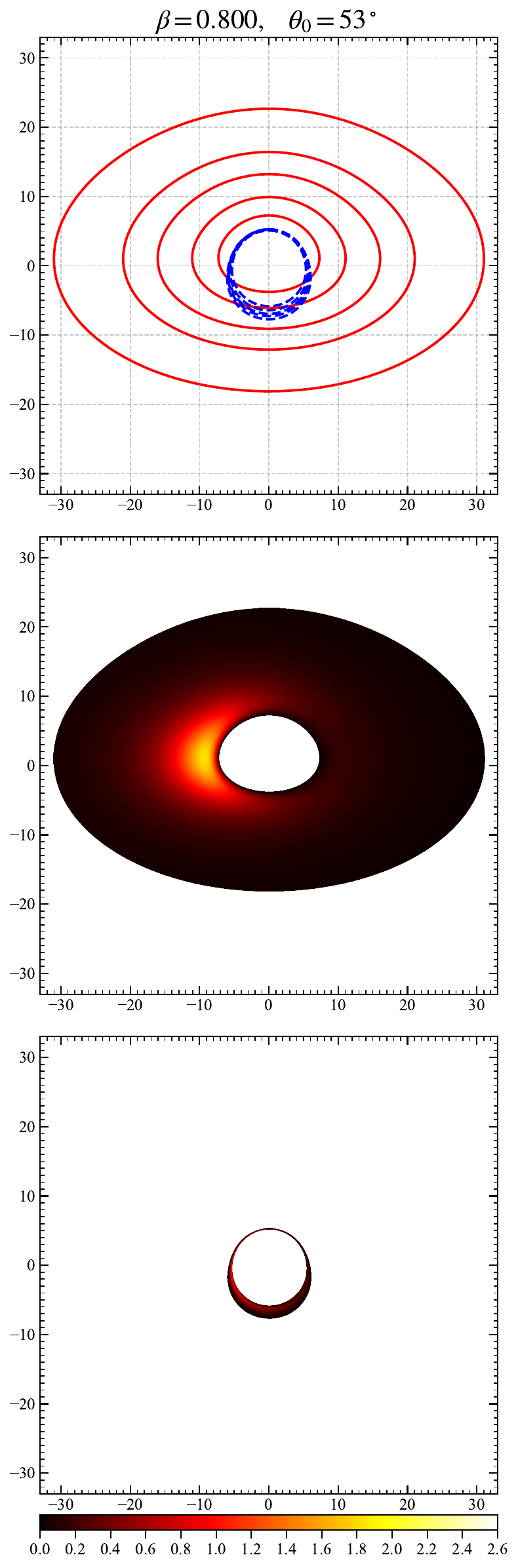}
    \end{subfigure}
    \hfill
    \begin{subfigure}{0.19\textwidth}
        \centering
        \includegraphics[width=\linewidth]{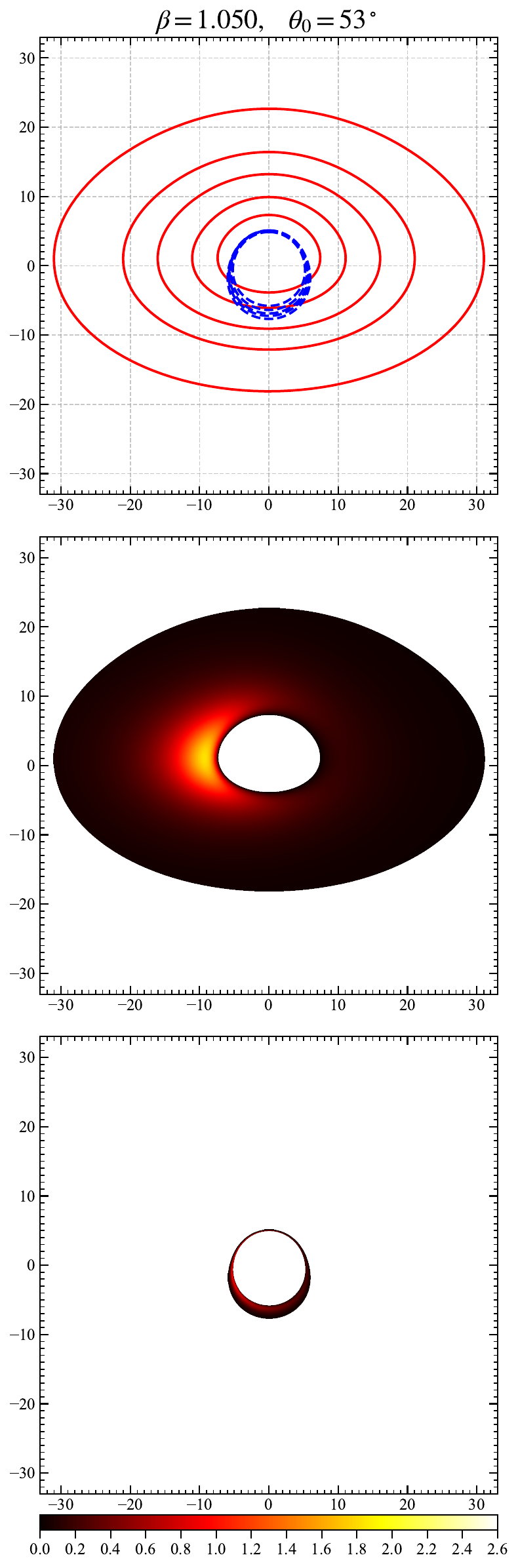}
    \end{subfigure}
    \hfill
    \begin{subfigure}{0.19\textwidth}
        \centering
        \includegraphics[width=\linewidth]{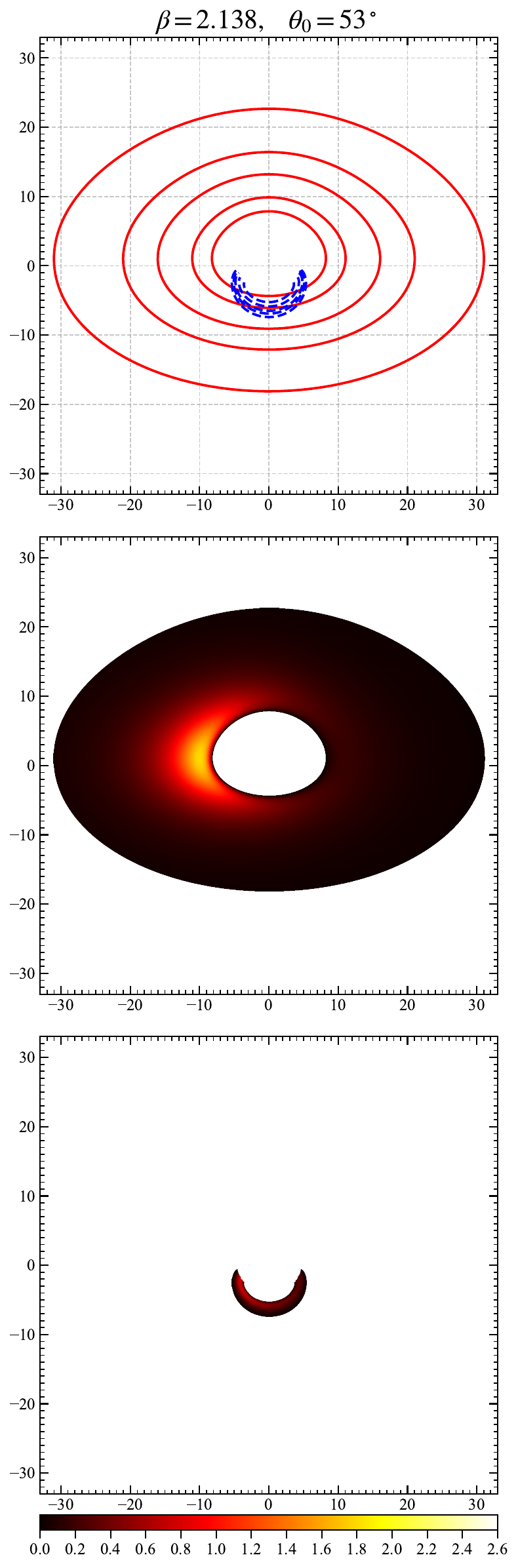}
    \end{subfigure}
    \begin{subfigure}{0.19\textwidth}
        \centering
        \includegraphics[width=\linewidth]{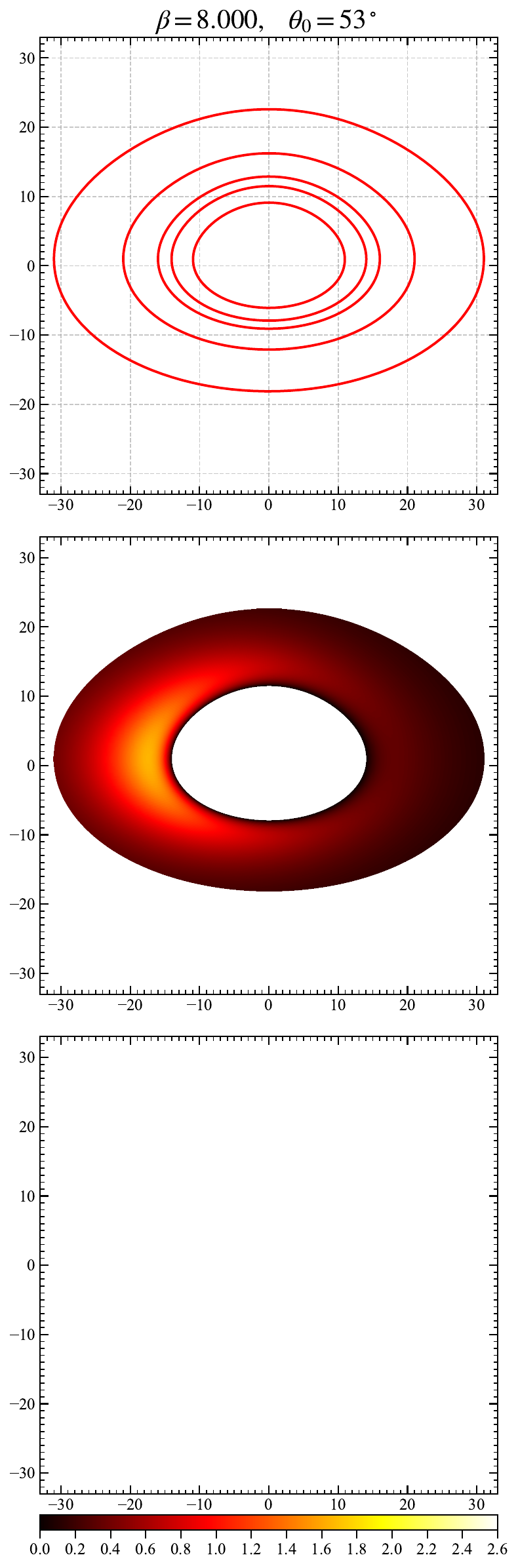}
    \end{subfigure}
    \caption{Images of geometrically thin and optically thick accretion disks for different values of $\beta$. The first row shows the projections of light rays emitted from sources at different radial positions onto the observer plane, corresponding from inner to outer radii $r_+, 10M, 15M, 20M,$ and $30M$, respectively. The red curves denote the primary image rays, while the blue curves denote the secondary image rays. The second row presents the observed intensity of the primary image, and the third row presents the observed intensity of the secondary image. For $\theta_0 = 53^\circ$, the secondary image gradually becomes incomplete and eventually disappears as $\beta$ increases.}  
    \label{fig:thick_image}  
\end{figure*}


From Fig.~\ref{fig:thick_image}, one can see that the direct image in this model is very similar to that of a black hole. This is because, outside the star, the gravitational properties are very similar to those of the spacetime outside a black hole event horizon, and thus the resulting accretion disk appears similar. However, because a regular star has neither an event horizon nor a singularity, its accretion disk image exhibits features absent in black hole spacetimes. The most prominent difference appears in the secondary images: when $\beta$ is sufficiently large, the secondary image may exhibit partial loss and ultimately vanish. This feature, which differs from black hole images, may provide a possible observational signature for future astronomers to identify compact objects composed of dark matter.

\section{Conclusions and discussion}


The study of non-singular black holes and black hole shadow is of current interest \cite{Regularspacetime1, Azreg1, Azreg2, Azreg3, Tsukamoto1, Tsukamoto2, Regularspacetime5, Aanna1, Aanna2} . In this work, we investigate the images of accretion disks around regular stars. We find that, in such spacetimes, photons can pass through the stellar center, giving rise to a central bright spot. This result is consistent with recent astronomical observations \cite{LensEinstein}. We study representative accretion disk models in this spacetime and find that, in addition to the transparency of the stellar center, the secondary images of the accretion disk become incomplete when the observation angle is sufficiently large. This feature may serve as a decisive observational signature for future astronomical observations aimed at identifying such regular stars.


In fact, regular spacetimes, including regular black holes and regular stars, constitute a special class of spacetimes and may arise from various mechanisms \cite{Regularspacetime1,Regularspacetime2,Regularspacetime3,Regularspacetime4,Regularspacetime5}. The most well-known approach derives regular metrics from nonlinear electrodynamics with magnetic charge, which implies that the matter content supporting such spacetimes behaves as a form of hot dark matter. However, it is also possible that the same metrics can be obtained from other types of fields. Given that the current standard cosmological model requires cold dark matter, we plan to focus in future work on identifying and studying the spacetime properties of regular stars supported by cold dark matter. We believe that such investigations will be valuable both for astronomical searches for dark matter and for theoretical studies of its fundamental properties.


\begin{acknowledgments}
The authors acknowledge funding support from the following agencies: FAPESP, grant 2014/07885-0 (EA); CNPq, grants 303592/2020-6 (EA); FAPESQ-PB; the National Natural Science Foundation of China (Grants: 42230207, 42074191, 12505060); the Shandong Provincial Natural Science Foundation, grant ZR2024QA032; Youth Innovation Group Plan of Shandong Province, grant 2023KJ107; the Fund Project of Chongqing Normal University (Grant Number: 24XLB033); Chongqing Natural Science Foundation General Program (Grant No. CSTB2025NSCQ-GPX1019).
\end{acknowledgments}


\end{document}